\documentstyle[11pt]{article}

\let\a=\alpha

 \def\bd{\begin{document}} \def\ed{\end{document}}
\def\ds{\documentstyle} \let\fr=\frac \let\bl=\bigl \let\br=\bigr
\let\Br=\Bigr \let\Bl=\Bigl 
\let\bm=\bibitem
\let\na=\nabla
\let\pa=\partial \let\ov=\overline 
\newcommand{\be}{\begin{equation}} 
\newcommand{\ee}{\end{equation}} 
\def\ba{\begin{array}}
\def\ea{\end{array}}
\def\ft#1#2{{\textstyle{{\scriptstyle #1}\over {\scriptstyle #2}}}}
\def\fft#1#2{{#1 \over #2}}
\def\del{\partial}
\def\vp{\varphi}
\def\sst#1{{\scriptscriptstyle #1}}
\def\oneone{\rlap 1\mkern4mu{\rm l}}
\def\td{\tilde}
\def\wtd{\widetilde}
\newcommand{\ho}[1]{$\, ^{#1}$}
\newcommand{\hoch}[1]{$\, ^{#1}$}
\newcommand{\bea}{\begin{eqnarray}} 
\newcommand{\eea}{\end{eqnarray}} 
\newcommand{\ra}{\rightarrow}
\newcommand{\lra}{\longrightarrow}
\newcommand{\Lra}{\Leftrightarrow}
\newcommand{\ap}{\alpha^\prime}
\newcommand{\bp}{\tilde \beta^\prime}
\newcommand{\tr}{{\rm tr} }
\newcommand{\Tr}{{\rm Tr} } 
\newcommand{\NP}{Nucl. Phys. }
\newcommand{\tamphys}{\it Center for Theoretical Physics\\
Texas A\&M University, College Station, Texas 77843}
\newcommand{\auth}{H. L\"u\hoch{\dagger},
S. Mukherji\hoch{\ddagger}, C. N. Pope\hoch{\dagger} and
J. Rahmfeld}

\begin{document}

\hfill{CTP TAMU-16/96}

\hfill{hep-th/9604127}

\vspace{20pt}

\begin{center}
{ \large {\bf Loop-corrected Entropy of Near-extremal Dilatonic $p$-branes}}

\vspace{30pt}

\auth

\vspace{15pt}

{\tamphys}

\vspace{40pt}

\underline{ABSTRACT}
\end{center}

     It has recently been shown that for certain classical non-dilatonic
$p$-branes, the entropy and temperature satisfy the ideal-gas relation
$S\sim T^{p}$ in the near-extremal regime.  We extend these results to cases
where the dilaton is non-vanishing, but nevertheless remains finite on the
horizon in the extremal limit, showing that the ideal-gas relation is again 
satisfied. At the classical level, however, this relation does break down if
the dilaton diverges on the horizon. We argue that such a divergence
indicates the breakdown of the validity of the classical approximation, and
that by taking string and worldsheet loop corrections into account, the
validity of the entropy/temperature relation may be extended to include
these cases. This opens up the possibility of giving a microscopic
interpretation of the entropy for all near-extremal $p$-branes.

{\vfill\leftline{}\vfill
\vskip	10pt
\footnoterule
{\footnotesize
     \hoch{\dagger}	Research supported in part by DOE 
Grant DE-FG05-91-ER40633 \vskip	-12pt} 	
\vskip	10pt {\footnotesize \hoch{\ddagger}	
      Research supported in part by NSF Grant	PHY-9411543
\vskip	-12pt}}

\pagebreak
\setcounter{page}{1}

     Semi-classical investigations have revealed many intriguing
thermodynamic properties of black holes \cite{h1,h2,b}.  They have also
raised many questions, related to the issues of predictability and
information loss in the quantum evolution of the system.  The resolution of
these questions is expected to require a more complete understanding of the
quantum theory of gravity.  It is widely believed that string theory will
supply this missing link.  Accordingly, in recent studies, various authors
have applied the semi-classical methods to the calculation of thermodynamic
quantities for black hole and higher extended-object solitons in string
theory or M-theory, and uncovered a striking relation to the microscopic
description of entropy in terms of the counting of string states [4--17].
A crucial ingredient in establishing this correspondence was that the
dilatonic scalar fields of the $p$-brane soliton remain finite on the
horizon in the extremal limit, in order that the tree-level approximation be
trustworthy.  Isotropic extremal $p$-brane solutions of this type are very
limited.  There are in total six cases, namely the elementary membrane
\cite{ds} and solitonic 5-brane \cite{g} in $D=11$, the self-dual 3-brane in
$D=10$ \cite{hs,dl2}, the dyonic string in $D=6$ \cite{dfkr}, and black
holes in $D=5$ and $D=4$ with three and four independent participating
field strengths respectively \cite{lpmult}, as well as a special dyonic
black hole in $D=4$, which involves two participating field strengths, 
each of which carries both electric and magnetic charges \cite{lpsol}. In
other words, in each case the number $N$ of non-vanishing charges compatible
with having some preserved supersymmetry must be maximal, namely
$N=1,1,2,3,4$ and 4 in $D=11,10,6,5,4$ and 4 respectively.  The general
non-extremal generalisations of these solitons can be found in \cite{dlp}.  

     In $D=11$, there is no dilaton field, and the microscopic
interpretation of the entropy of the membrane and 5-brane was presented in
\cite{kt}.  The dilaton field decouples from the self-dual 3-brane in $D=10$,
and the microscopic discussion of the entropy was presented in \cite{gkp}.
In the remaining three cases, the dilatonic scalar fields do not decouple
for generic values of the charges, but they do remain finite at the horizon
in the extremal limit. When all the charges are equal, the dilatonic scalar
fields then decouple; the resulting solutions were referred to as 
``non-dilatonic'' $p$-branes in \cite{kt}. In this case, the dyonic string
becomes the self-dual string in $D=6$, and the black holes become
Reissner-Nordstr\o{}m black holes in $D=5$ and $D=4$ respectively. (The
dilaton field also decouples for the special dyonic black hole in $D=4$,
when the electric and the magnetic charges are equal.) In fact it is
sometimes the case that multiply-charged black holes \cite{cy} may be
regarded as bound states at threshold of singly-charged black holes
\cite{dr,sen,dlr,r}.  All the above solitons are regarded as
``regular-dilaton'' $p$-branes.  (In contrast, ``dilatonic'' solutions in
general refer to $p$-branes whose dilatonic scalar fields do not decouple
for any non-vanishing charges; moreover, these scalar fields diverge at the
horizon in the extremal limit.) The $D=5$ black hole solution is stainless
\cite{lpss}, but can be oxidised to a boosted dyonic string in $D=6$
dimensions. The entropy per unit $p$-volume is preserved under the
oxidisation, and was shown to be precisely equal to that of the microscopic
counting of the corresponding D-string states in the near-extremal regime
\cite{sv}.  There are two regular black holes in $D=4$.  One of them
involves 4 field strengths, and reduces to the Reissner-Nordstr\o{m} black
hole when all the charges are equal.  The analogous analysis has also been
carried out for this black hole \cite{jkm,sm}.  The other involves only two
field strengths, each with electric and magnetic charges.  This solution was
referred to as the dyonic black hole of the second type in \cite{lpsol}.  We
shall extend the previous entropy analysis to include this case. 

      The first five cases above are the only $p$-brane solitons whose entropies
have so far been given a microscopic interpretation.  Furthermore, it has
been shown that in the non-dilatonic cases, the entropy and temperature
satisfy the ideal-gas relation $S\sim T^p$ \cite{kt}, which is in
concordance with the microscopic D-brane picture.  We shall show that this
dicussion can be extended to the regular dilatonic cases. 
There are, however, many more $p$-brane solitons in the string or M-theory,
in addition to the six cases discussed above.  At the classical level, the
ideal-gas entropy/temperature relation breaks down in all these other cases.
This can be attributed to the fact that for all these solitons the dilaton,
as well as the curvature and field strengths, diverges on the horizon in the
extremal limit. These divergences indicate a breakdown of validity of the
classical approximation, and it can be argued that the inclusion of string
and worldsheet loop corrections will remove such singularities.  We present
a simple explicit example that  illustrates how this happens.  Then, we
study the relation between entropy and temperature in the presence of loop
corrections, and show that the ideal-gas entropy/temperature relation may be
recovered.  This  may open up the possibility of giving a microscopic
interpretation of the  entropy for all near-extremal $p$-branes. 

      Let us begin with a discussion of the thermodynamic properties of
classical $p$-branes.  The relevant part of the bosonic Lagrangian is the
tree-level approximation of the string effective action, namely 
\bea
e^{-1}{\cal L} = R -\ft12 (\del \vec \phi)^2 -\fft1{2n!} \sum_{\a=1}^N 
e^{-\vec a_\a \cdot \vec \phi} F_\a^2\ ,\label{treelag}
\eea
where $\vec \phi=(\phi_1,\ldots, \phi_{\sst N})$, and $F_\a$ is a set of $N$
$n$'th-rank antisymmetric tensor field strengths, which give rise to a 
$p$-brane with world volume dimension $d=n-1$ if they carry electric
charges, or with $d=D-n-1$ if they carry magnetic charges.  The ``dilaton
vectors'' $\vec a_\a$ are constant vectors characteristic of the supergravity
theory arising as the low energy limit of the string or the M-theory.  
Note that the Lagrangian (\ref{treelag}) can be embedded into the string
or the M-theory if the dot products $M_{\a\beta} = \vec a_\a\cdot \vec
a_\beta$ satisfy \cite{lpmult} 
\be
M_{\a\beta} = 4\delta_{\a\beta} - \fft{2d \td d}{2(D-2)}\ ,\label{mmatrix}
\ee
where $\td d = D-d-2$.  

     The metric of the black $p$-branes with $N$ non-vanishing charges is 
given by \cite{dlp}
\bea
ds^2 &=& e^{2A} (-e^{2f} dt^2 + dx^i dx^i) + e^{2B}
(e^{-2f} dr^2 + r^2 d\Omega^2)\ ,\label{metricform}\\
e^{2A}&=&\prod_{\a =1}^{N} \Big( 1 + \fft{k}{r^{\td d}} \sinh^2 \mu_\a 
\Big)^{-\ft{\td d}{D-2}} \ ,\qquad e^{2B}=\prod_{\a=1}^N \Big( 1 +
\fft{k}{r^{\td d}}\sinh^2\mu_\a\Big)^{\ft{d}{D-2}} \ ,\label{blackmetric} 
\eea
where the coordinates $(t, x^i)$ parameterise the $d$-dimensional 
world-volume of the $p$-brane, and the remaining coordinates of the $D$ 
dimensional spacetime are $r$ and the coordinates on the 
$(D-d-1)$-dimensional unit sphere, whose metric is $d\Omega^2$.  The 
function $f$ has a completely universal form \cite{dlp}:
\be
e^{2f} = 1 -\fft{k}{r^{\td d}}\ .\label{fsol}
\ee
The dilatonic scalar fields $\varphi_\a = \vec a\cdot \vec\phi$ are given by
\be
e^{-\ft12 \epsilon\varphi_\a} = (1 + \fft{k}{r^{\td d}}\sinh^2\mu_\a) 
\prod_{\beta=1}^{N} \Big( 1 + \fft{k}{r^{\td d}} 
\sinh^2\mu_\beta\Big)^{-\ft{d\td
d}{2(D-2)}}\ , \label{dilaton} 
\ee
where $\epsilon=1$ for elementary solutions and $\epsilon=-1$ for solitonic
solutions. The metric (\ref{blackmetric}) has an outer horizon at
$r=r_+\equiv k^{1/\td d}$ and a curvature singularity at $r = r_-\equiv 0$.
In general the curvature remains singular at the origin in the extremal
limit $r_+ \rightarrow r_-=0$.  The mass per unit $p$-volume and the charges
for each field strength are given by 
\be
m = k(\td d + 1) + k\td d\sum_{\a =1}^{N}\sinh^2\mu_\a\ ,\qquad
\lambda_\a = \ft12 \td d k \sinh 2\mu_\a\ .\label{seven}
\ee
The extremal limit corresponds to sending $k\rightarrow 0$ and $\mu_\a 
\rightarrow \infty$ while keeping the charges $\lambda_\a$ fixed.  When all
the charges are equal, the solutions reduce to single-scalar $p$-branes,
described by the Lagrangian 
\be
e^{-1}{\cal L} = R -\ft12(\del \phi)^2 -\fft{1}{2n!} e^{-a\phi} F^2\ ,
\label{slag}
\ee
where the constant $a$ can be parameterised by $a^2 = \Delta -2d\td d/(D-2)$, 
with $a$ given by $a^2=(\sum_{\a,\beta} (M^{-1})_{\a\beta})^{-1}$
\cite{lpsol}, implying that $\Delta = N/4$.  In this case, the functions $A$
and $B$ in the metric are given by \cite{dlp} 
\be
e^{2A} = \Big(1 + \fft{k}{r^{\td d}} \sinh^2\mu\Big)^{-\ft{4\td 
d}{\Delta(D-2)}}\ ,\qquad
e^{2B} = \Big(1 + \fft{k}{r^{\td d}} \sinh^2\mu\Big)^{\ft{4
d}{\Delta(D-2)}}\ .\label{singlescal}
\ee
Note that for choices of the dot products $M_{\a\beta}$ other than those
given by (\ref{mmatrix}), the equations of motion following 
from (\ref{treelag})  can be cast into the form
of Toda-like equations \cite{lpx}, which have not yet been solved.  However
their single-scalar truncations are solvable, yielding solutions of the same 
form (\ref{singlescal}), with  values of $\Delta$ other than $4/N$.  These
solutions are not supersymmetric in the extremal limit. 

     The temperature and the entropy per unit $p$-volume for the $p$-brane 
metric (\ref{blackmetric}) are given by \cite{dlp}
\be
T = \fft{\td d}{4\pi r_+} \prod_{\a=1}^N (\cosh\mu_\a)^{-1}\ ,\qquad
S = \ft14 r_{+}^{\td d +1}\, 
\omega_{\td d+1} \prod_{\a=1}^{N} \cosh \mu_\a\ ,
\ee
where $\omega_{\td d +1} = 2\pi^{\td d/2 +1}/(\ft12 \td d)!$ is the volume 
of the unit $(\td d+1)$-sphere.  For brevity we shall often refer to $S$,
which is the entropy per unit $p$-volume, simply as the entropy. In the
near-extremal regime,  {\it i.e.}\ $k<<\lambda_\a$ for all
$\a$,\footnote{The term ``near-extremal regime'' is used  rather loosely
here since in fact we are requiring the stronger condition  that $k$ be much
smaller than each of the individual charges, rather than  merely the sum of
all the charges, so that we can approximate  $\sinh 2\mu_\a$ in (\ref{seven})
by $\ft12 e^{2\mu_\a}$.} the temperature and the entropy are therefore
related by 
\be
S \approx\gamma \Big(\prod_{\a=1}^{N} \lambda_\a\Big)^{\ft{\td d}{N\td d -2}}
\,\, T^{\ft{2(\td d+1) - N\td d}{N \td d-2}} \ ,
\ee
where $\gamma = \td d \omega_{\td d+1} (16\pi^2 \td d^{-N-2})^{\td d/(N\td 
d-2)}/(16\pi)$.  When the number $N$ of charges in the $p$-brane solution
satisfies 
\be 
N=2(D-2)/(d\td d)\ ,\label{ncondition}
\ee
we find that the entropy/temperature relation becomes
\be
S \approx \gamma  \Big(\prod_{\a=1}^{N} \lambda_\a\Big)^{\ft{d}{2}}
\,\, T^{d -1}\sim T^p\ ,\label{etr}
\ee
which has precisely the natural massless ideal-gas scaling, predicted by
D-brane considerations since open strings on a Dirichlet $p$-brane can be
viewed as an ideal gas of massless objects in a $p$-dimensional space. 
Note that if the charge parameters $\lambda_\a$ are all equal, in which 
case, the dilatonic scalars decouple, the ideal-gas entropy/temperatur 
relation (\ref{etr}) reduces to the one found in \cite{kt}.  It is
interesting that this entropy/temperature relation holds even when the
charges are not equal, in which case the dilatonic scalar fields do not
decouple, although they do remain finite at the horizon in the extremal
limit.  Note that the temperature goes to zero in the extremal limit for all
$p$-brane solitons satisfying the condition (\ref{ncondition}).  On the
other hand, such $p$-brane solitons have vanishing entropy in the extremal
limit only for $p\ge1$. Note also  that when the condition
(\ref{ncondition}) is satisfied, the entropy of the near-extremal $p$-branes
can also be expressed as 
\be
S\sim (\delta m^2)^{\ft{d-1}d}\ ,\label{density}
\ee
which is consistent with the asymptotic density of states of $p$-branes
\cite{ao}. Thus another approach that can yield a microscopic 
interpretation of the entropy is by counting the states of a fundamental 
$p$-brane.

     It follows from (\ref{dilaton}) that if (\ref{ncondition}) is
satisfied, the dilatonic scalar fields
$\varphi_\a$ are finite at the horizon in the extremal limit.  Moreover, we
have $\sum_\a\varphi_\a=0$, and hence there are a total of $(N-1)$
non-vanishing scalar fields.  It is worth remarking that the curvature and
the field strengths are also finite at the horizon in the extremal limit
when the condition (\ref{ncondition}) is satisfied.  There are precisely
five cases where $N$ satisfies (\ref{ncondition}).    When $N=1$, this
corresponds to the elementary membrane or the solitonic 5-brane in $D=11$
where there is no dilaton field, or to the self-dual 3-brane in $D=10$,
where the dilaton decouples. The $N=2$ case corresponds to the dyonic string
in $D=6$, with one dilaton field.  The $N=3$ and 4 cases correspond to the
black holes in $D=5$ and $D=4$ with 2 and 3 dilatonic scalar fields
respectively.  In the latter three cases, the dilatonic scalar fields all
decouple when the charges $\lambda_\a$ are all set equal. 

     For other values of $N$, (or, in the case of single-scalar solutions,
other values of $\Delta$) the entropy/temperature relation $S\sim T^{d -1}$
in the near-extremal regime breaks down.  This breakdown of the relation can
be expected since in these cases, the physical quantities such as dilatonic
scalar fields, the field strengths and the curvature diverge at the horizon
in the extremal limit.   This indicates that for dilatonic $p$-branes, the
tree-level approximation is not sufficient.  In fact, as we shall argue
next, the divergence of these physical quantities at the horizon in the
extremal limit may merely be an artifact of the tree-level approximation to
the effective action of the string.  If we include also 
loop effects, the dilaton field, as well as the field strengths and the
curvature,  may be finite at the horizon, and the entropy/temperature
relation of the resulting $p$-branes will satisfy precisely the natural
massless ideal-gas scaling.  In other words, these loop corrections
have the effect of turning the tree-level singular-dilaton $p$-branes into
regular-dilaton $p$-branes. 

     The inclusion of string and worldsheet loop corrections is in general
very  complicated; they will include higher powers of curvature and field 
strengths, and different exponents in the effective dilaton couplings
\cite{dl}. Later, we shall argue that subject to rather general assumptions
about the effect of higher-loop corrections, the ideal-gas
entropy/temperature relation will be restored.  First, let us consider a
simple example which illustrates the phenomenon, namely the
heterotic string in $D=6$, obtained by compactification of the $D=10$ 
heterotic string on $K3$.  In particular let us consider black hole 
solutions whose charges are carried by the Yang-Mills fields.  The string
loop-corrections in the effective Lagrangian have been calculated 
\cite{sag}, and the relevant part is given by
\be
e^{-1} {\cal L} = R - \ft12(\del\phi)^2 -\fft{1}{2n!} g(\phi) F^2\ ,
\label{slagloop}
\ee
where
\be
g(\phi) = v e^{-\phi/\sqrt2} + \td v e^{\phi/\sqrt2}\ ,\label{gphi}
\ee
and $v$ and $\td v$ are constants.  The first term in (\ref{gphi}) is the 
classical contribution, and its coefficient $v$ is always positive
since it is essentially the Kac-Moody level \cite{e,dm,dmw}.  The constant
$\td v$ can sometimes be negative. In this case, the kinetic energy of the
gauge field becomes negative in the strong-coupling regime, indicating a
phase-transition \cite{dmw,sw,dlp2}. We shall restrict our attention in this
paper to the cases where $\td v$ is positive and hence the phase transition
does not occur. The function $g(\phi)$ satisfies 
\be
\fft{\del g(\phi)}{\del \phi}\Big|_{\phi=\phi_0} = 0\ \label{phicon}
\ee
for $e^{\sqrt2\phi_0} = v/\td v$.  This property, which is obviously not
satisfied by the tree-level term alone, will play an important role in
constructing regular-dilaton $p$-branes. 

    Given the standard $p$-brane ansatz for the field strength, black hole
solutions for the loop-corrected Lagrangian (\ref{slagloop}) can be easily
constructed.  In fact, we can generalise the discussion to Lagrangians of
the form (\ref{slagloop}) where $F$ represents a field strength of arbitrary
degree with some dilaton coupling of the form $g(\phi) = e^{-a\phi} +
\cdots$, where the first term is the classical contribution.  Let us suppose
that, as in the explicit example of the $D=6$ string discussed above, the 
function $g(\phi)$ satisfies (\ref{phicon}) for some $\phi_0$.  It is then 
easy to see that there is a $p$-brane solution where the dilaton is the
constant $\phi=\phi_0$, and the metric is given by 
\bea
ds^2 &=& e^{2A} (-e^{2f} dt^2 + dx^i dx^i) + e^{2B}
(e^{-2f} dr^2 + r^2 d\Omega^2)\ ,\nonumber\\
e^{-dA}&=& 1 + \fft{k}{r^{\td d}} \sinh^2 \mu 
\ ,\qquad e^{\td d B}=1 +
\fft{k}{r^{\td d}}\sinh^2\mu\ ,\label{loopmetric} 
\eea
with $f$ again given by (\ref{fsol}).
The mass per unit volume and the charge are given by
\be
m= \fft{2(D-2)k}{d} \sinh^2\mu + k(\td d+1)\ ,\qquad
\lambda = k\sqrt{\fft{(D-2)\td d}{2dg(\phi_0)}} \sinh2\mu\ .
\ee
As for the previous black $p$-branes, the metric (\ref{loopmetric}) has an
outer horizon at $r_+ = k^{1/\td d}$, and a curvature singularity at $r=0$. 
The  temperature and the entropy per unit $p$-volume are given by
\be
T = \fft{1}{4\pi r_+} (\cosh\mu)^{-\ft{2(D-2)}{d\td d}}\ ,\qquad
S = \fft14 r_+^{\td d+1} \omega_{\td d+1} (\cosh\mu)^{\ft{2(D-2)}{d\td d}}
\ ,\label{ts2}
\ee
The extremal limit $r_+\rightarrow 0$ is achieved by sending 
$k\rightarrow 0$ and $\mu \rightarrow \infty$ while keeping the charge
$\lambda$ fixed.  In the  near-extremal regime, the functions $A$ and $B$
behave as
\be
e^{A} \sim r_{+}^{\td d/d}\ ,\qquad e^{B} \sim \fft{1}{r_{+}}\ 
\ee
at the horizon.  All the physical observables, including the field strength,
are finite in this limit.  In particular the curvature at $r=r_{+}$, which
in the tree-level approximation was divergent as $r_+\rightarrow 0$, is now,
owing to the inclusion of loop corrections, regular in the extremal limit. 
It follows from (\ref{ts2}) that the ideal-gas entropy/temperature relation
(\ref{etr}) is restored.  Note  that in this case the entropy and energy
$\delta m$ also satisfy the  relation (\ref{density}). 

       The explicit $D=6$ example above was rather exceptional, in that the
2-index field strength that we used for constructing the black-hole solution
was derived from the Yang-Mills sector of the theory, rather than from the
fields of the supergravity multiplet.  Thus although a black-hole solution
with an extremal limit can be constructed, it will not preserve any
supersymmetry in this limit.  Furthermore, the form (\ref{slagloop}) of the
loop corrections is especially simple here, with no powers of $F$ higher
than the quadratic order arising.  Owing to this simplicity, we were able to
find an exact solution, and to show how the string-loop corrections gave
rise to modifications that led again to the ideal-gas form of the
entropy/temperature relation that was previously seen only for the
regular-dilaton $p$-branes.  We now turn to a general discussion of the
expected form of the quantum-corrected $p$-brane solutions, to show that
under rather broad assumptions, the quantum corrections can be expected to
restore the ideal-gas entropy/temperature relation. 

     It has often been argued that the higher-order quantum corrections to 
the string effective action can be expected to smooth out the singularities 
that may arise in solutions of the purely classical limit.  In particular, 
it is reasonable to expect that the quantum corrections will make all
the physical quantities in the $p$-brane solutions finite at the outer 
horizon. In the example above, this phenomenon was indeed observed; in the 
string-loop corrected theory, the dilaton remained finite on the horizon (in
fact, it  became constant everywhere), and furthermore all invariant
quantities built  from the fields, such  the curvature scalar and the square
of the field  strength, $F_{\sst{MN}} F^{{\sst {MN}}}$, remained finite on
the horizon.  In other examples, it may be that worldsheet loop
corrections, rather than string loops, ensure finiteness of the physical
quantities in the extremal limit. This is reminiscent of the
stretched-horizon approach \cite{sen}.

     To proceed in general, let us make the assumption that the effect of
loop corrections on {\it any} elementary or solitonic $p$-brane solution
will be to render all physical invariant quantities finite at the horizon in
the extremal limit. Thus we may consider such solutions that still have the
general isotropic form (\ref{metricform}), with the usual form of isotropic
$n$-index field-strength configuration, 
\be
F = \lambda \, *\epsilon\ ,\qquad{\rm or}\qquad F=\lambda\, \epsilon\ ,
\label{fansatz}
\ee
where $\epsilon$ is the volume form on the unit $(D-d-1)$-sphere.  
Owing to the loop corrections however, the precise form of the solution will
no longer be given by (\ref{blackmetric}). If we calculate the square of the 
field strength, using (\ref{fansatz}) and (\ref{metricform}), we find that 
it is proportional to $(r\, e^{B})^{-2n}$, and thus to be finite on the 
horizon in the extremal limit, we must have
\be
r_+\, e^{B(r_+)}\longrightarrow  c \ ,\label{finite}
\ee
where $c$ is some constant.  Furthermore, we find that finiteness of the 
curvature scalar for  the metric (\ref{metricform}) at $r=r_+$ again requires 
(\ref{finite}) in the extremal limit, together with some further regularity
condition for the function $A$.  The entropy per unit $p$-volume, and the
temperature, are given by \cite{dlp} 
\bea
S=\ft14 r_+^{\td d+1} \, \omega_{\td d+1}\, 
e^{(\td d+1) B(r_+) + (d-1) A(r_+)}\ ,\label{entrop}\\
T=\fft{\td d}{4\pi r_+}\, e^{A(r_+)-B(r_+)}\ .\label{temp}
\eea
Substituting (\ref{finite}) into these expressions, we 
therefore find that the entropy and temperature for the loop-corrected 
$p$-brane will take the form
\be
S=\ft14 c^{\sst D-2}\, \omega_{\td d+1}\, (\fft{4\pi}{\td d})^{d-1}\, 
T^{d-1}\ 
\ee
in the near-extremal regime. Thus we see that the assumption of regularity of
physical fields on the  horizon even in the extremal limit implies that the
entropy/temperature relation will always take the ideal-gas form.  Note that 
for the loop corrected $p$-branes satisfying the above ideal-gas relation, 
the explicit form of the function $A$ is unimportant.

      To summarise, we have studied the relation between entropy and
temperature for near-extremal $p$-branes.  It was recently shown \cite{kt}
that in classical $p$-brane solitons where the dilaton is either absent (as
in $D=11$) or constant, the entropy is proportional to $T^p$.  This relation
is the one that is found for an ideal gas of massless particles in $p$
spatial dimensions, which is in line with expectations from D-brane
state-counting arguments.  We have extended this result to include cases 
where the dilaton does not decouple, provided that it is regular on the 
horizon in the extremal limit.  The entropy/temperature relation breaks
down in cases where the dilaton is singular on the horizon in the extremal
limit.  In fact the majority of classical $p$-brane solitons exhibit such
singularities, and singularities in the curvature and field strengths,
indicating a breakdown of the tree-level approximation.  It can be argued
that the inclusion of loop corrections or non-perturbative effects will
remove these divergences. We have demonstrated that this assumption is
sufficient to recover the ideal gas entropy/temperature relation for all the
near-extremal $p$-branes, and hence may remove the previous obstacles to
providing a microscopic interpretation of the entropy by the counting of
states.

\end{document}